\documentclass{iaus}
\usepackage{graphicx}
\usepackage{psfig}
\newcommand{\figstart}[1]  {\begin{figure} \psfig{#1}}
\newcommand{\lfigstart}[1] {\begin{figure*} \psfig{#1}}
\newcommand{\figend} {\end{figure}}
\newcommand{\lfigend} {\end{figure*}}

  \checkfont{eurm10}
  \iffontfound
    \IfFileExists{upmath.sty}
      {\typeout{^^JFound AMS Euler Roman fonts on the system,
                   using the 'upmath' package.^^J}%
       \usepackage{upmath}}
      {\typeout{^^JFound AMS Euler Roman fonts on the system, but you
                   dont seem to have the}%
       \typeout{'upmath' package installed. iaus.cls can take advantage
                 of these fonts,^^Jif you use 'upmath' package.^^J}%
      }
  \else
  \fi

  \checkfont{msam10}
  \iffontfound
    \IfFileExists{amssymb.sty}
      {\typeout{^^JFound AMS Symbol fonts on the system, using the
                'amssymb' package.^^J}%
       \usepackage{amssymb}%

      }{}
  \fi

  \IfFileExists{amsbsy.sty}
    {\typeout{^^JFound the 'amsbsy' package on the system, using it.^^J}%
     \usepackage{amsbsy}}
    {\providecommand\boldsymbol[1]{\mbox{\boldmath $##1$}}}

\newsavebox{\astrutbox}
\sbox{\astrutbox}{\rule[-5pt]{0pt}{20pt}}

\newcommand\etal{\mbox{\textit{et al.}}}

\newcommand\Sauron{{\tt SAURON}}

\title[The Interplay among Black Holes, Stars and ISM in Galactic 
       Nuclei]{Why (inner) bars are important but not sufficient}

\author[E. Emsellem]%
{Eric Emsellem$^1$%
}

\affiliation{$^1$Centre de Recherche Astronomique de Lyon, Saint Genis Laval, France \\
email: emsellem@obs.univ-lyon1.fr }

\pubyear{2004}
\volume{222}
\pagerange{1--8}
\date{?? and in revised form ??}
\jname{The Interplay among Black Holes, Stars and ISM \\in Galactic Nuclei}
\editors{Th. Storchi Bergmann, L.C. Ho \& H.R. Schmitt, eds.}
\begin{document}

\maketitle

\begin{abstract}
I briefly report on the work conducted to probe the gravitational
potential of active and non-active disk galaxies using gas and
stellar kinematics.
\end{abstract}

\firstsection 
\section{Introduction}

Bars and non-axisymmetric galactic structures imply time variable torques which
can help transfering angular momentum. Large-scale bars are important
actors in the spatial redistribution of the dissipative component in disk 
galaxies, and there is indeed a direct observed link between their presence
and e.g. star formation or the gas concentration. However, as repeatedly
emphasized by many authors (see e.g. Martini's contribution in these Proceedings),
there is no significant correlation between the nuclear activity and the presence
of such structures. A nice summary has been given by F. Combes 
(these Proceedings), but I should start by repeating a simple but important argument.

\section{Time and space}

Before discussing the causality between different physical phenomena, it is
first necessary to clearly define what we wish to write about, and
to examine if the presumed actors can coexist in time and space.
When astronomers write about fuelling a galactic nucleus, they
mean the central engine, where most of the high energy radiation originates: 
close to the central black hole, and well inside a few hundredths of parsec.
It is therefore not that surprising if we cannot find direct evidence for a link between
such a tiny region and large (few kpc) or medium (few hundred pc) scale structures.

Time is then also a critical parameter. If we just list a few important processes
and their corresponding timescales:
\begin{itemize}
\item Dynamics: 100~pc corresponds to about $10^6$~yr for a velocity of 100 km/s.
\item The duty cycle of nuclear activity is usually quoted with scales of $10^6$ to $10^8$~yr.
\item Steady mass accretion:  $10^3$ to $10^4$~yr (see Wada, these Proceedings). 
\item Average fuelling rate: $10^{-3}$ to 1~M$_{\odot}$/yr
(for a total mass processed of $10^3$ to $10^8$~M$_{\odot}$).
\item Star formation has a global timescale of about $10^8$~yr.
\end{itemize}
If the central activity and the formation of a large-scale bar are reccurent processes
(see Combes, these Proceedings), it seems illusory to expect an observed simultaneity
(Fig.~\ref{fig:arrow}.
But as gas is assumed to gradually fall down the gravitational potential of the galaxy,
we can proceed in steps, focusing on one specific spatial scale at a time. 
\figstart{figure=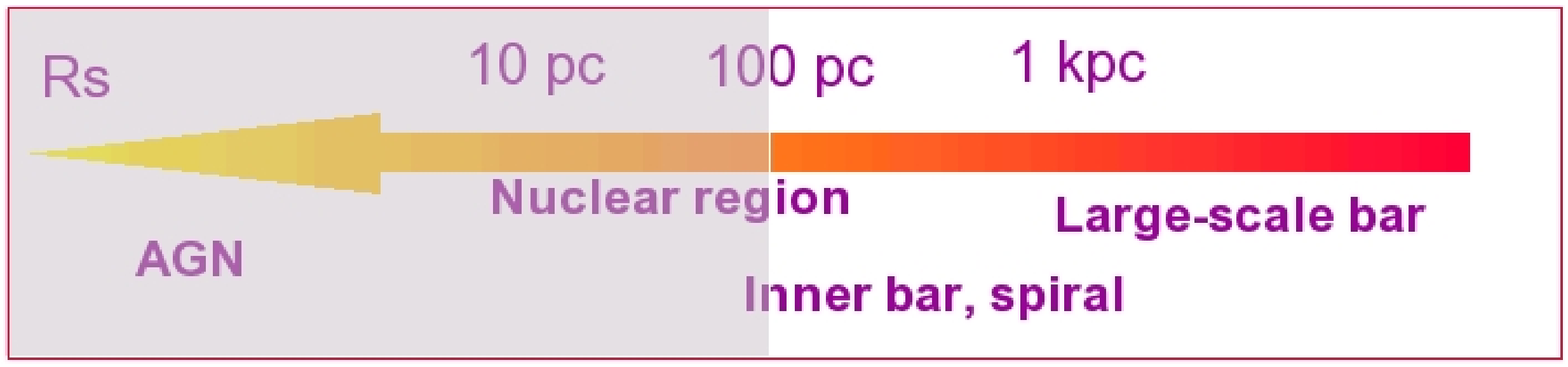,width=\textwidth}
\caption[]{Sketch showing the wildly different scales we usually deal with when
trying to link the nuclear activity with e.g. a large-scale bar. Rs is the Schwarzschild
radius of the presumed black hole, $\sim 10^{-5}$~pc for $10^8$~M$_{\odot}$. }
\label{fig:arrow}
\figend

\section{Building a gas reservoir at 1 kpc}

The first important step can be taken to be roughly 1 kpc, as it corresponds
to e.g. the Inner Lindblad Resonance of large-scale bars where gas can
be efficiently accumulated. Although the loss of angular momentum
is intrinsically a dynamical process, we do not know much
about the underlying gravitational potential of disk galaxies. 

We thus defined a small sample of nearby early-type disk galaxies with the aim
of constraining both their stellar and gas distributions and kinematics.
A range of very nearby galaxies with matched active and non-active pairs
was carefully selected. For each target we are probing sensitive large-scale
tracers, using HI observations, but also more centrally concentrated ones
such as the ionized gas and the stellar populations within the central few kpc.
The gas and stellar kinematics are derived via integral-field spectroscopy with
\Sauron, a large field unit mounted on the William Herschel telescope. 

We expect a number of difficulties to pave our way. 
We will need to disentangle the gas and stellar contributions in spectra
which contain strong and wide emission lines. We also need to interpret
the gas and stellar kinematics in a consistent framework using adaptive 
modelling techniques. NGC~1068 was observed in Jan. 2002 with \Sauron\ 
and used as a technical benchmark in this
context. We thus developed an iterative method to properly extract the kinematic information,
using synthetic stellar library and a penalized pixel fitting routine 
to fit the stellar absorption lines (see Cappellari \& Emsellem 2004).
We also built N body/SPH simulations starting from axisymmetric initial conditions
to probe the effect of the observed near-infrared bar on the stellar kinematics.
We could thus provide some new constraints on the dynamics, and time variability
of the system. Work is in progress and will be published soon (Emsellem \etal, in preparation).

We obtained our first \Sauron\ observations of active and non-active galaxies
in March 2004: the weather has been favourable enough so we could observe
12 targets (6 pairs). The reduction and analysis will be conducted using
our dedicated pipeline. The survey will be completed with two runs in the 2004B and 2005A
semesters. 

\section{Building a gas reservoir at 100 pc}

Zooming towards the center, at a scale of about 100~pc, we are still quite
far from the direct influence of the presumed supermassive black hole:
for a typical stellar velocity dispersion of 150~km/s, the radius
of influence is only $r_h = 2 (M_{\bullet} / 10^7~M_{\odot})$~pc.
In order to build a gas reservoir inside $\sim 100$~pc (not invoking
mergers), we could think of inner density waves such as secondary
bars, central spirals or $m=1$ modes.

Inner bars are present in at least 25\% of all barred galaxies (Erwin et al. 04, 
Erwin and Sparke 2002), and this represents more than 15\% of all disk galaxies.
Another 20\%  of galaxies contain inner disks, which may also
be related to fuelling episodes (see e.g. van den Bosch \& Emsellem 1998). 
Secondary bar sizes range from 200 to 800~pc (Erwin et al. 04)
which means that the corresponding (presumed) Inner Lindblad Resonances should 
be well within the central 100~pc.  

The above-mentioned fraction of galaxies containing innner bars 
is of course a lower limit as such structures are not easily detected.
Studies have so far focused on photometric and dust features, 
but our specific study of NGC~2974 has shown that these could
be misleading (Emsellem, Goudfrooij, Ferruit, 2003), hence 
weak bars could easily be hidden by the central spheroidal component.
We believe that the best tracers include the stellar and gas distribution
and kinematics, and that again only their detailed mapping can reveal
the nature of the underlying potential.

This is clear when looking at cases like NGC~1358, NGC~3504 with
an integral-field spectrograph such as OASIS. The two-dimensional
gas distribution and kinematics shows highly non-circular motions
with the inner bar and/or $m=1$ structures leading to significant
inward motion. Unfortunately, it is sometimes difficult (or impossible)
to disentangle the contribution of the AGN (outflow, nuclear emission)
from the gravitational motion. 

\section{The formation and evolution of $\sigma$--drops}

This is nicely illustrated in
the case of NGC~5728, where OASIS observations
show bar-driven kinematics on one side of the nucleus, and
strong AGN-driven outflow on the other side.
In this case, we need to probe the stellar component which is expected
to be less prone to non-gravitational motions. We obtained VLT/ISAAC
long-slit data in the central few arcseconds of NGC~5728 and derived
the stellar kinematics along the major and minor axes of the inner bar
(Emsellem et al. 2001). These revealed a clear velocity decoupling of the central arcsecond,
($\sim 150~pc$). More interestingly,
the stellar velocity dispersion profile exhibits a significant drop in the
same region. These so-called $\sigma$--drop were in fact observed in 3 out of 3
galaxies for which we could probe the central kinematics.
$\sigma$--drops are now almost routinely observed in disk galaxies
(e.g. Marquez et al. 2003), and a non-exhaustive list is provided in
Fig.~~\ref{fig:sdrop}. It is important to emphazise that
such drops require high signal-to-noise spectra with reasonable 
spatial resolution, and that data of this type
are scarce for disk galaxies.
\figstart{figure=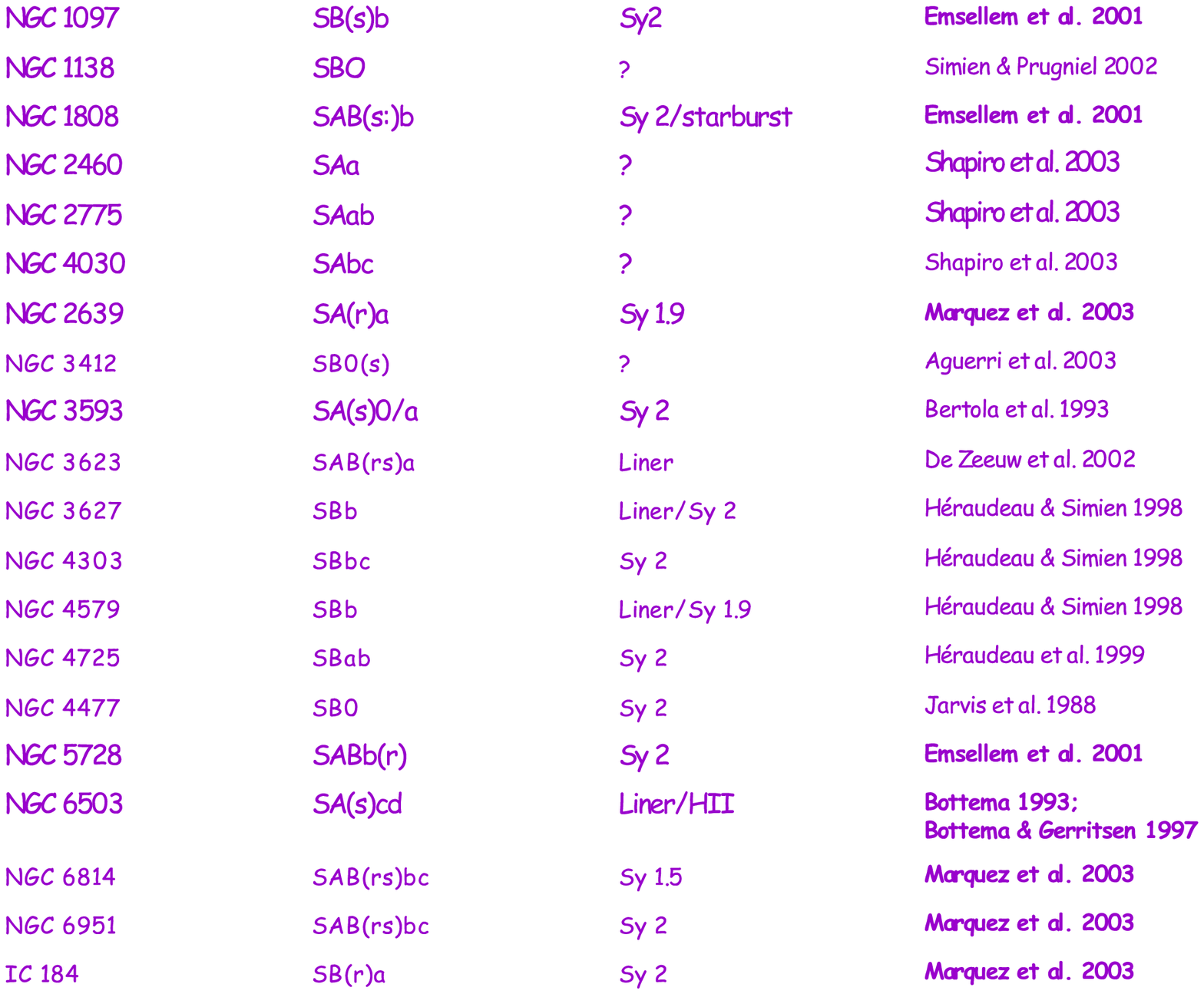,width=\textwidth}
\caption[]{Non-exhaustive list of galaxies with observed $\sigma$--drops.
From left to right: galaxy name, type, activity if known, and associated reference.}
\label{fig:sdrop}
\figend

We have studied a possible scenario for the formation of $\sigma$--drops
via N body $+$ SPH simulations including star formation (Wozniak et al. 2003). 
A decrease in the stellar velocity dispersion requires the presence
of a dynamically cold component, which we assume is formed via bar-driven
gas infall and subsequent star formation. This scenario is nicely supported by
the numerical simulations which exhibits long-lived $\sigma$--drops.
The new-born stars form out of very low dispersion gas, and the $\sigma$--drop
appears where the young stellar disk significantly contributes to the luminosity.
The whole central stellar system, not just the disk, will heat up with time, 
allowing the $\sigma$--drop to stay visible for hundreds of Myr.
We should emphasize that this is however not a unique scenario, 
and any process which can efficiently
tranport gas within the central few hundreds parcsec is a good candidate
for the formation of $\sigma$--drops. 

\section{Conclusions}\label{sec:concl}

I will conclude by reiterating on a few important issues. First stating
once more that building a gas reservoir within the central 1~kpc or 100~pc,
is a different thing than fuelling the central engine: spatial and time scales
are important in this context. Hence we should make it clear when we 
write about ''fuelling the AGN''. 

In this short paper, I focused on the accumulation of gas first within
the central 1000 and 100~pc. Our \Sauron\ survey of active and non-active
galaxies, supplemented by large-scale HI data, 
will provide us with a unique probe of the underlying gravitational potential
in the inner kpc. Inside this region, inner bars are not rare and may indeed
play an important role in accumulating mass within the central 100~pc.
However, we need to map both gas and stellar kinematics in order to reveal
the dynamical structures. We may then find that $\sigma$--drops are
long-lived signatures of accretion, but this will require to extend
the available pool of data quite dramatically.

\begin{acknowledgments}
I wish here to thank my main collaborators in these projects:
the \Sauron\ team, Fran\c{c}oise Combes, Kambiz Fathi. Pierre Ferruit, 
Daniel Friedli, Bruno Jungwiert, Carole Mundell, Neil Nagar, Herv\'e Wozniak.
\end{acknowledgments}

\end{document}